\documentclass[preprint]{revtex4}
\usepackage[utf8]{inputenc}
\usepackage{CJKutf8}
\usepackage{graphicx}
\usepackage{amsmath}
\usepackage[dvipsnames]{xcolor}
\usepackage[colorlinks=true, allcolors=blue]{hyperref}
\usepackage[a4paper, total={6in, 8in}]{geometry}
\usepackage{float}
\usepackage[parfill]{parskip}
\usepackage{dcolumn}

\begin{document}
\begin{CJK*}{UTF8}{}

\title{Nanosecond True Random Number Generation with Superparamagnetic Tunnel Junctions - Identification of Joule Heating and Spin-Transfer-Torque effects}
\author{Leo Schnitzspan}
\author{Mathias Kläui}
\author{Gerhard Jakob}
\affiliation{Institute of Physics, Johannes Gutenberg-University Mainz, 55122 Mainz, Germany}
\affiliation{Max Planck Graduate Center Mainz, 55122 Mainz, Germany}

\maketitle
\end{CJK*}

\section*{Abstract}
This work investigates nanosecond superparamagnetic switching in $50\,$nm diameter in-plane magnetized magnetic tunnel junctions (MTJs).
Due to the small in-plane uniaxial anisotropy, dwell times below $10\,$ns and auto-correlation times down to $5\,$ns are measured for circular superparamagnetic tunnel junctions (SMTJs).
SMTJs exhibit probabilistic switching of the magnetic free layer, which can be used for the generation of true random numbers.
The quality of random bitstreams, generated by our SMTJ, is evaluated with a statistical test suite (NIST STS, sp 800-22) and shows true randomness after three XOR operations of four random SMTJ bitstreams.
A low footprint CMOS circuit is proposed for fast and energy-efficient random number generation.
We demonstrate that the probability of a 1 or 0 can be tuned by spin-transfer-torque (STT), while the average bit generation rate is mainly affected by the current density via Joule heating.
Although both effects are always present in MTJs, Joule heating most often is neglected.
However, with a resistance area (RA) product of $15\,\Omega$µm$^2$ and current densities of the order of $1\,$MA/cm$^2$, an increasing temperature at the tunneling site results in a significant increase in the switching rate.
As Joule heating and STT scale differently with current density, device design can be optimized based on our findings.

\section{Introduction} 

A superparamagnetic tunnel junction (SMTJ), acting as a random stochastic noise source, is the counterpart to non-volatile random access memory (MRAM)  \cite{MRAM_overview1}, where high temporal stability (switching energy barrier $E_b>40\,k_{\mathrm{B}}T$) of the state is required to store the information of a bit. 
The volatile behavior of the superparamagnetic junction originates from the ambient thermal energy acting on the magnetization of the free layer.
This thermal energy is high enough to overcome the low energy barrier of a few $k_{\mathrm{B}}T$, resulting in a superparamagnetic state with fluctuation times of milliseconds down to nanoseconds at room temperature.
The time spent in each state, called dwell time, can be controlled by an applied external magnetic field, by spin-transfer-torques, or spin-orbit torques \cite{SOT_SMTJ1}.
Due to the inherent probabilistic nature of SMTJs, state controllability, and energy-efficiency, SMTJs were proposed for various computational concepts \cite{neuro_spintronics_rev1}, such as invertible logic  \cite{Camsari_pbits}, Boltzmann machines \cite{BM_pbits}, reservoir computing \cite{RC_pbits}, spiking neural networks \cite{spiking_NN1, spiking_NN2} or stochastic computing \cite{SC_Daniels}. 
These concepts can provide advantages over pure digital complementary metal-oxide-semiconductor (CMOS) based computational logic, since the fundamental building block is a stochastic bit.
Despite the tremendous development of conventional CMOS-based deterministic computers over the last decades, there are still classes of problems that can not be addressed efficiently, due to the deterministic nature of von Neumann computers.
Many numerical computation techniques are based on the Markov Chain Monte Carlo (MCMC) methods \cite{MCMC_book} and require many random numbers.
Since SMTJs are inherently probabilistic, as opposed to deterministic CMOS circuits, they can provide random signals with low-power and low areal footprint, which can be transformed to random bitstreams using a few CMOS transistors.
MTJs are already CMOS compatible on scale for different applications such as MRAM \cite{MRAM_overview1}.
Through the combination of CMOS logic and a set of probabilistic SMTJs (called p-bits \cite{Camsari_pbits}), it has been shown that computationally hard problems like the "traveling salesman problem" can be solved efficiently \cite{Camsari_traveling_salesman}.
The performance of such a "p-computer" will then be a consequence of the p-bit hardware density and the average fluctuation rate, described by the Néel-Arrhenius law. 
Fast fluctuations and dwell times in the nanosecond range are desired.
For in-plane easy-axis MTJs (ip-MTJs), dwell times of the order of milliseconds \cite{ip_SMTJ_ms_1}, microseconds \cite{ip_SMTJ_us_1, ip_SMTJ_us_Majetich, ip_SMTJ_us_Majetich2} down to nanoseconds \cite{ip_SMTJ_ns_Sun, ip_SMTJ_ns_Fukami} have been reported.
Compared to out-of-plane MTJs (oop-MTJ), ip-MTJs often exhibit shorter dwell times, due to a different contribution of in-plane and out-of-plane anisotropy energies \cite{theory_relaxation_times}.
Out-of-plane MTJs often comprise larger time scales of milliseconds \cite{oop_SMTJ_ms_Reiss, oop_SMTJ_ms_Majetich} to microseconds \cite{Camsari_inter_fac, oop_SMTJ_us_Majetich, oop_SMTJ_us_Fukami} and for this reason, we have decided to design our MTJ stack with in-plane ferromagnetic layers.
We demonstrate that two phenomena occur when a current is applied to an MTJ nanopillar:
spin-transfer-torque (STT) and Joule heating. 
STT refers to the transfer of angular momentum from the flowing electrons to the magnetic moments in the ferromagnetic layer. 
Joule heating, on the other hand, refers to the generation of heat induced by the flow of charge current through the device. 
Both effects are always present in an MTJ and have to be considered to understand the dependence of fluctuation times and current density.
So far, Joule heating and STT have been studied together on superparamagnetic nanoislands with spin-polarized scanning tunneling microscopy \cite{Joule_heating_Krause}, but Joule heating has mostly been ignored or neglected for superparamagnetic tunnel junctions, where only STT effects were considered.\\
In this work, we separate and extract the contributions of STT and Joule heating for different current densities to understand the mechanism of the superparamagnetic switching dependence.
We show that STT affects the state energy linearly with applied current, while Joule heating has a quadratic dependence on the current.
Due to the low energy barrier between both states, Joule heating results in an overall faster switching, while STT tunes the state probability differently.
In case of a random number generator, it is important to control the state probability to a desired value.
An implementation idea based on logic XOR gates is proposed in order to generate a stream of true random bits.
The quality of randomness is quantified by the statistical test suite from NIST \cite{NIST_STS}.
We show that true randomness, with a bit generation rate of $200\,$Mbit/s, is achieved by using four SMTJs and a combination of three XOR gates.
It is also demonstrated that for the generation of multiple independent true random bitstreams, two XOR gates and two SMTJs are sufficient.
This circuit design saves energy and space and therefore has the potential for applications where many random bits are required.

\section{Methods} 
\subsection{MTJ sample preparation}

\begin{figure}[H]
\centering
\includegraphics[width=0.9\textwidth]{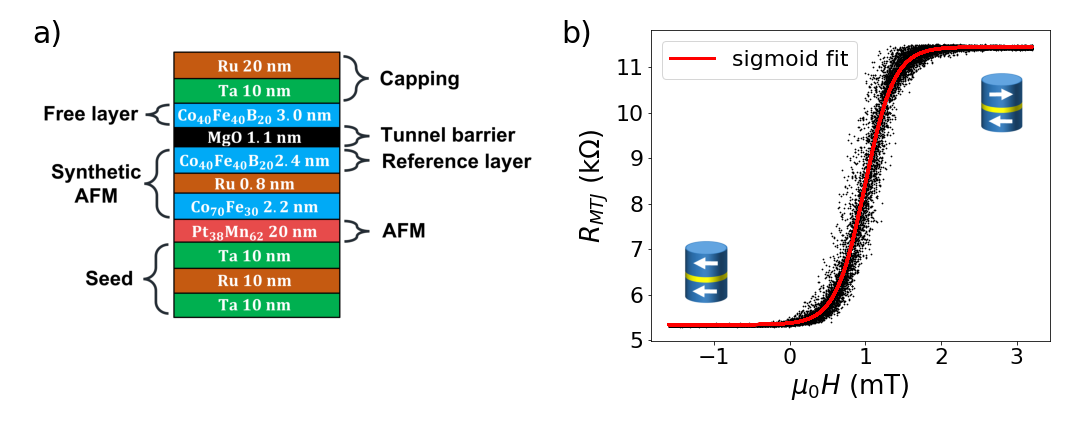}
\caption{\label{fig:RH_loops}a) TMR stack with RA-product of $15\,\Omega$µm$^2$ and TMR ratio of $150\,\%$ used for SMTJs. b) MTJ resistance versus external applied easy-axis field of a superparamagnetic tunnel junction is plotted for 20 repetitions. The red line indicates the sigmoidal relation of resistance and magnetic field. Negative fields stabilize the parallel state, while positive fields stabilize the antiparallel state.}
\end{figure}

TMR stacks were deposited at room temperature on oxidized Si substrates using rf- and dc-magnetron sputtering (Singulus Rotaris) with the following composition (film thickness in nanometer): Ta(10)/Ru(10)/Ta(10)/PtMn(20)/CoFe(2.2)/Ru(0.8)/CoFeB(2.4)/MgO(1.1)/CoFeB
(3.0)/Ta(10)/Ru(20) and is based on an optimized stack, developed earlier \cite{LS_ADI_paper}.
The stack is illustrated in Fig.\ \ref{fig:RH_loops}a, where the function of each layer is specified.
The tunnel magnetoresistance (TMR) ratio of our stack is found to be approximately $150\,\%$ and for patterned nanopillars, it is in the range of $100 - 150\,\%$.\\
The state of the MTJ (parallel or antiparallel) is determined by the magnetization orientation of the free layer with respect to the reference layer.
The CoFeB reference layer together with a CoFe pinned layer forms a synthetic antiferromagnet (SAF), in order to compensate stray fields at the free layer site.
In addition, the pinned CoFe layer is exchange biased \cite{theory_exchange_bias} by a PtMn antiferromagnet (AFM).
Even though a SAF is integrated into the stack, the stray field at the free layer position will never be compensated exactly to zero \cite{strayfield_pMTJ}.
For this reason, an in-plane easy axis field of a few mT is typically applied in order to compensate for this "offset" and to set the SMTJ into equal state probability. 
This offset is approximately $1.1\,$mT in Fig.\ \ref{fig:RH_loops}b but has a large variation from device to device due to variations in the sample fabrication process.
As shown in Fig.\ \ref{fig:RH_loops}b, the state probability can be tuned by the applied in-plane easy axis field.
A change in the field strength results in a shift of the energy levels, thus stabilizing either the P- or AP-state.
The relative increase in resistance from the P- to the AP-state is defined as the TMR ratio: $TMR = R_{AP}/R_{P}-1$. 
To obtain a high TMR ratio, annealing was carried out in a $300\,$mT in-plane field at $300\,^{\circ}$C for $1\,$h.
Current in-plane tunneling (CIPT) measurements \cite{cipt_tech} of our unpatterned stack exhibit a resistance area (RA) product of $15 \pm 2 \,\Omega$µm$^2$ and a TMR ratio of $150\pm3\,\%$.
MTJ nanopillars are patterned using 30\,kV electron-beam lithography, to define a circular etch mask. Argon ion etching under 55° and 20° is used to structure the magnetic tunnel junction pillar.
As a passivation layer, SiN$_x$ was chosen for hardness and good insulating properties.
It was sputtered under a 50/50 mixed argon-nitrogen atmosphere at $5\,$Pa.
Above the MTJ nanopillar, a $60\,$nm thick gold pad was sputtered as a top electrode lead in order to be able to bond it to a measurement sample holder.
By using the top electrode and the conducting seed layer as a bottom electrode, a voltage across the junction was applied and the switching behavior was studied. \\

\subsection{Set-up and measurement}
Electrical measurements were carried out at room temperature with an MTJ in series with a shunt resistor $R_s$. 
With a $3.5\,$GHz bandwidth digital oscilloscope (Tektronix DPO7543, 40\,Gs/s), the amplified signal of the superparamagnetic tunnel junction fluctuation is measured.
A $50\, \Omega$ low noise amplifier (ZFL-1000LN+, mini-circuits) with a gain of $20\,$dB is used to amplify the signal.
To assure that the time series measurement has a high time resolution and is not limited by the RC time constant of the set-up, parasitic capacitance and resistances of the set-up have to be small.
The $50\,\Omega$ input impedance of the amplifier results in a small signal voltage drop of the order of $1\,$mV at the amplifier.
Since the MTJ resistor ($5-10\,$k$\Omega$) is in parallel with the $50\,\Omega$ input impedance of the amplifier, the equivalent resistance is approximately also $50\,\Omega$ (neglecting the impedance of $C_i$ for high frequencies).
Forming a voltage divider with $R_s$ leads to a small AC voltage under stochastic fluctuation of $R_{MTJ}$ (see appendix S1: Eq.\ \ref{eq:SMTJ_signal}).
In addition, a low resistance lowers the RC time constant of the set-up, thus leading to a high time resolution of approximately $1\,$ns ($R\,\approx50\,\Omega$ and $C\,\approx 20\,$pF). 
Therefore, time series of the fluctuating voltage signal are typically recorded with a sampling rate of 1\,GS/s for a few milliseconds.
The measurement set-up is illustrated in Fig.\ \ref{fig:time_series}a and the recorded nanosecond time series is plotted in Fig.\ \ref{fig:time_series}b.

\section{Dwell times and auto-correlation times}
\begin{figure}[H]
\centering
\includegraphics[width=0.99\textwidth]{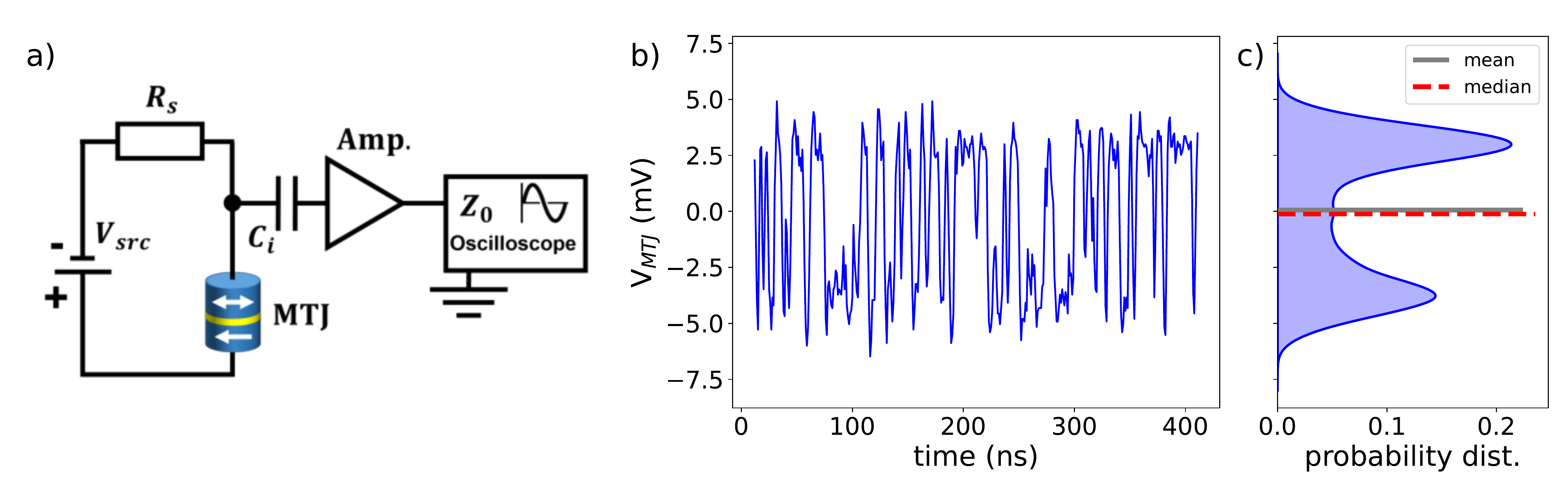}
\caption{\label{fig:time_series}a) Circuit diagram of the measurement set-up. b) Example of the nanosecond stochastic switching of the amplified SMTJ signal. c) Probability distribution (kernel density estimate) of the SMTJ signal for $10^5$ data points.}
\end{figure}
\begin{figure}[H]
\centering
\includegraphics[width=0.7\textwidth]{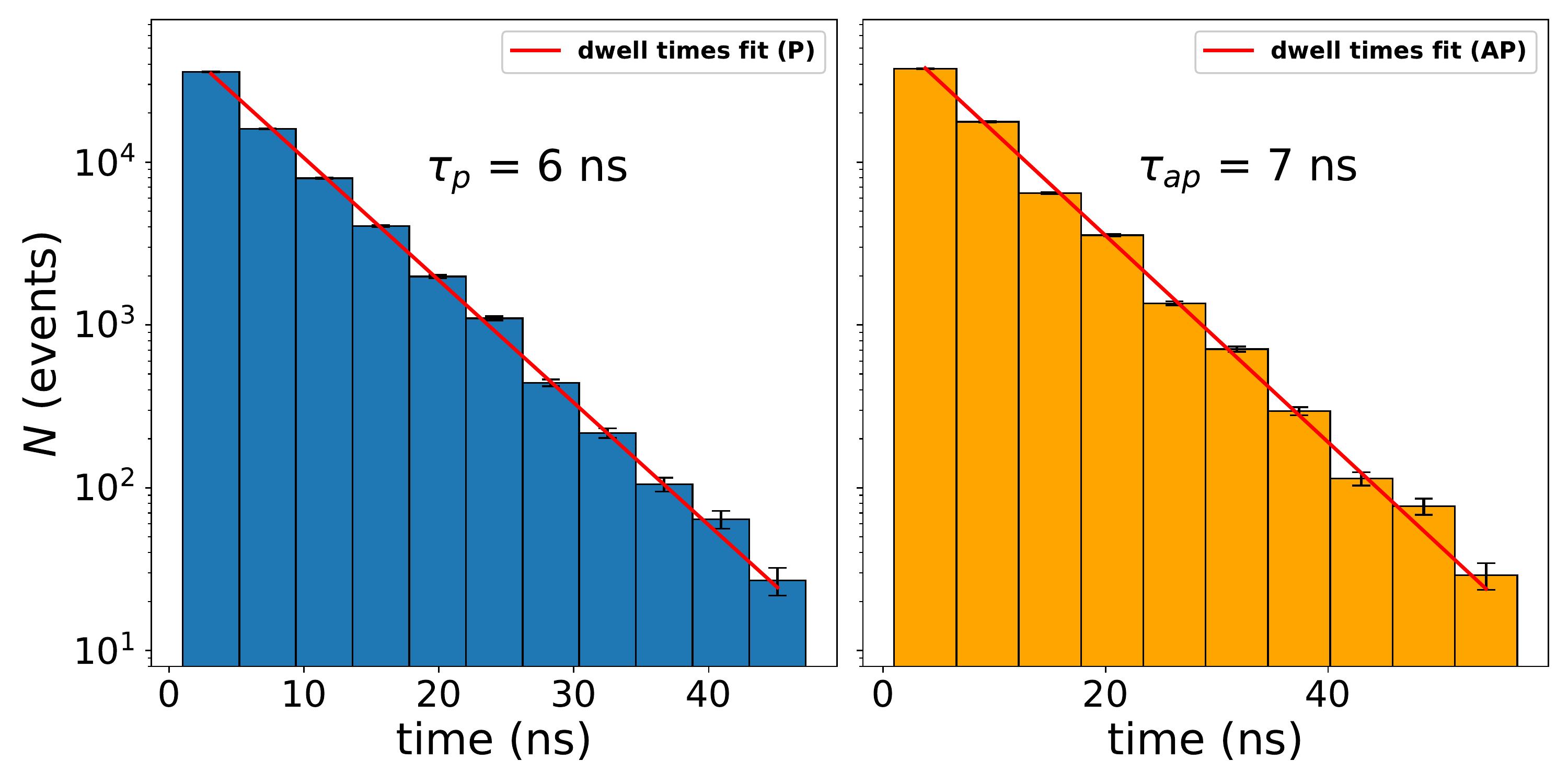}
\caption{\label{fig:poisson_plot}Histogram of dwell times for a $1\,$ms time series
for the P and AP state. Bins can be interpreted as a waiting time interval in a Poisson process for which exactly one switching event will occur in this interval. From a fit (red line) the average dwell times $\tau_p$ and $\tau_{ap}$ are determined, according to a Poisson process.}
\end{figure}
\begin{figure}[H]
\centering
\includegraphics[width=0.7\textwidth]{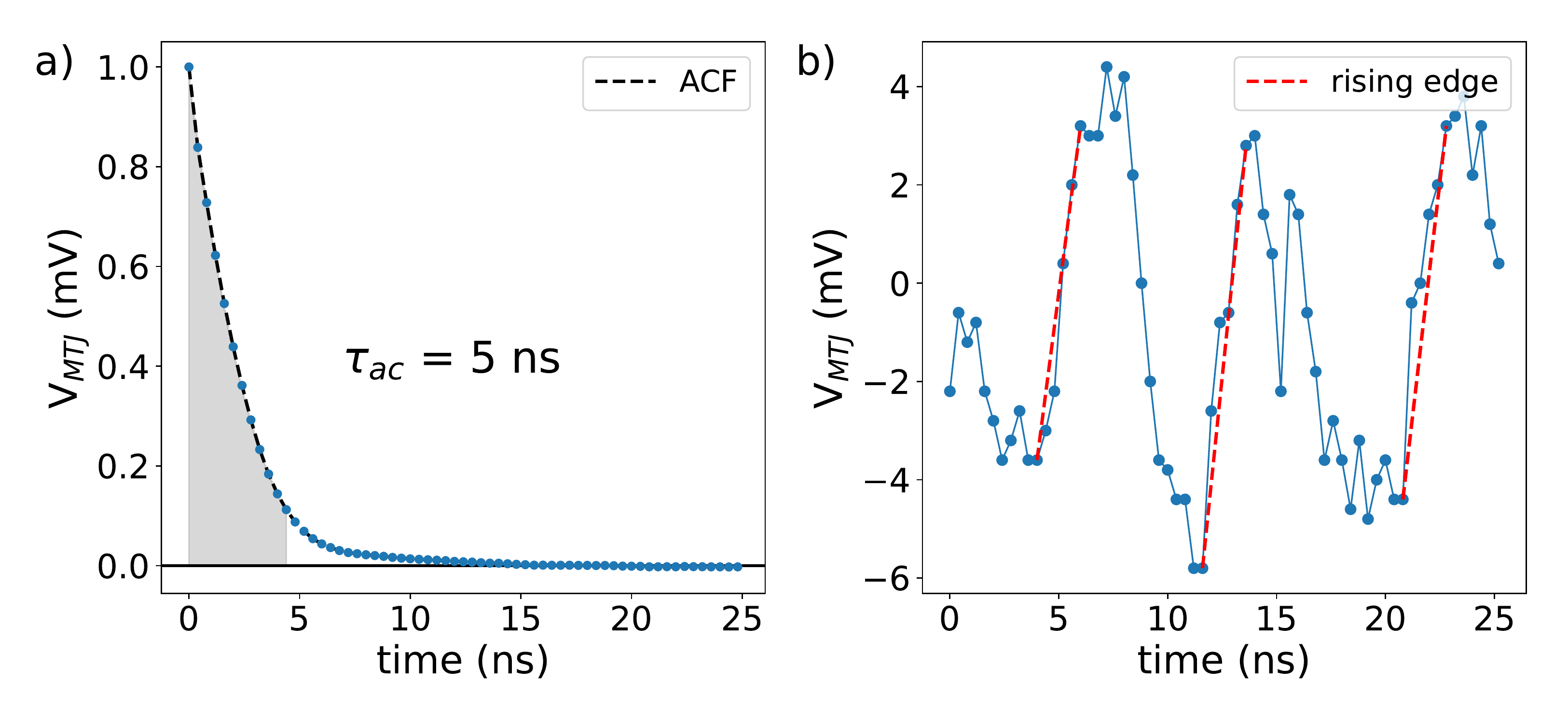}
\caption{\label{fig:ACF_and_rise_time}a) Auto-correlation function (ACF) is plotted for an SMTJ time series measurement. $\tau_{ac}$ is defined as the 99\% integral under the ACF curve. b) A detailed view of the rising and falling edges of the time series. Three red dashed lines show three rising edges, which are approximately linear and of the order of $2\,$ns.}
\end{figure}

A high random bit generation rate is primarily limited by the average fluctuation rate or dwell time.
The dwell time in the macrospin approximation for $E_b/k_{\mathrm{B}} T \geq 1$, follows the Néel-Arrhenius law \cite{theory_Neel_law}:
$\tau_{p,ap} = \tau_0 \, \exp \left( E_b/k_{\mathrm{B}} T \right)$,
where $E_b$ is the energy barrier between the states, $T$ the temperature and $\tau_0$ the attempt time.
In order to generate true random bits, the prerequisite is the independence of each generated bit with respect to the previously generated bit.
An important metric for this is the auto-correlation time $\tau_{ac}$ of the SMTJ signal, which can be interpreted as the time scale for randomness and describes for which time interval the signal is still correlated to its past.
As shown theoretically \cite{theory_low_barrier_m_IEEE}, MTJs with in-plane magnetic anisotropy exhibit shorter auto-correlation times than MTJs with out-of-plane anisotropy due to the difference in precession-like fluctuations, attributed to the large demagnetization field for ip-MTJs.
In the macrospin approximation with a single normalized magnetization vector $m=M/M_s$ fluctuating randomly on a Bloch sphere, the anisotropy energies will restrict the fluctuation to a specific region in the sphere. 
The demagnetization energy constrains the fluctuation to the equator of the Bloch sphere and the uniaxial anisotropy to regions near the easy axis directions.
In order to measure these superparamagnetic fluctuations, a circular nanopillar MTJ with $50\,$nm in diameter was patterned by electron-beam lithography and structured with Argon-ion etching.
With the measurement set-up shown in Fig.\ \ref{fig:time_series}a, stochastic fluctuations have been acquired, where the DC component of the signal is filtered out by the capacitance in the circuit.
The time between two consecutive switching events is defined as dwell time and is measured for the parallel as well as the anti-parallel state. 
Dwell times can be determined in two ways.
First, a time series measurement ($0.5\,$s sampling rate: $1\,$GS/s) is binarized, by using the median as a reference threshold, and then the series is divided into an array of AP (=1) and P (=0) dwell time intervals.
The mean of all AP and P time intervals then corresponds to the average dwell time for the AP and P state.
The average dwell time $\tau$ is calculated by $\sqrt{\tau_p \tau_{ap}}$, as suggested in literature \cite{ip_SMTJ_ns_Fukami}.
For the average dwell time, we find $\tau=7.3\pm0.2\,$ns.
Second, the switching mechanism can be described by a Poisson process, where the probability for a switching event follows the Poisson distribution, similarly as in a radioactive decay, where the decay probability is described by a Poisson process.
By fitting the number of switching events for the AP and P state with $N= N_0 \exp(-t/\tau_{(p,ap)})$, the average dwell times $\tau_p, \tau_{ap}$ can be extracted in a more robust manner, as shown in Fig.\ \ref{fig:poisson_plot}.
All calculated (average) dwell times of this work are determined by the second method through fitting a Poisson distribution to the dwell times.
Fig.\ \ref{fig:poisson_plot} indicates that the distribution of dwell times is in agreement with the theory, as shown by a linear decrease in the logarithmic plot.
The average dwell time is found to be $\tau = 6.7\pm0.1\,$ns, which is orders of magnitude shorter than most measured dwell times of oop-SMTJs or ip-SMTJs.
We further determine the auto-correlation time of the binarized signal, which is the limiting factor for the generation of random numbers.
The auto-correlation function (ACF) is calculated by the following equation, where $s$ is the binarized signal, $\overline{s_i}$ the mean of it and $t$ the time lag of the time series of length $N$:
\begin{equation}
ACF(t) = \frac{\sum_{i=1}^{N-t} [s_i - \overline{s_i}]  [s_{i+t} - \overline{s_i}]} {\sum_{i=1}^{N} (s_i - \overline{s_i})}
\end{equation}
The ACF is plotted in Fig.\ \ref{fig:ACF_and_rise_time}a and shows an exponential decrease.
After the auto-correlation time $\tau_{ac}$, the signal is uncorrelated to any signal of its past.
It is defined as the maximum time by which the integral of the ACF reaches $99\,$\%.
This results in an auto-correlation time of $\tau_{ac}=5.1\pm0.3\,$ns.
Next, we consider the rise time of the signal, which was determined by averaging 100 rising edges in a time series (sampling rate: $2.5\,$GHz).
Typically the definition of the rise time $\tau_{rise}$ is a 10\,\% to 90\,\% change of amplitude on the rising edge of a signal. 
However, this stochastic telegraph noise has a noise contribution, coming from the set-up, as well as a small stochastic variation in the signal for both P- and AP-states due to the thermal excitation of the magnetization vector.
For this reason, we define the rise time as the time between the midrange point minus one standard deviation to the midrange point plus one standard deviation: $\tau_{rise}=t_{mid+\sigma} - t_{mid-\sigma}$ The midrange point is defined as $1/2(\max(s)+\min(s))$. 
Of 100 rising edges the average rise time is determined to be $\tau_{rise}=1.5\pm0.6\,$ns, as shown in Fig.\ \ref{fig:ACF_and_rise_time}b.
The RC time constant has to be shorter than the rise time and is approximately of the order of $1\,$ns for $R\approx50\,\Omega$ and $C\approx20\,$pF.  
The rising edge (or falling edge) does not indicate an exponential increase (decrease), suggesting that the time resolution is high enough to acquire the signal and that it is not affected by the RC time constant of the set-up.

\section{True random number generation}
\begin{figure}[H]
\centering
\includegraphics[width=0.9\textwidth]{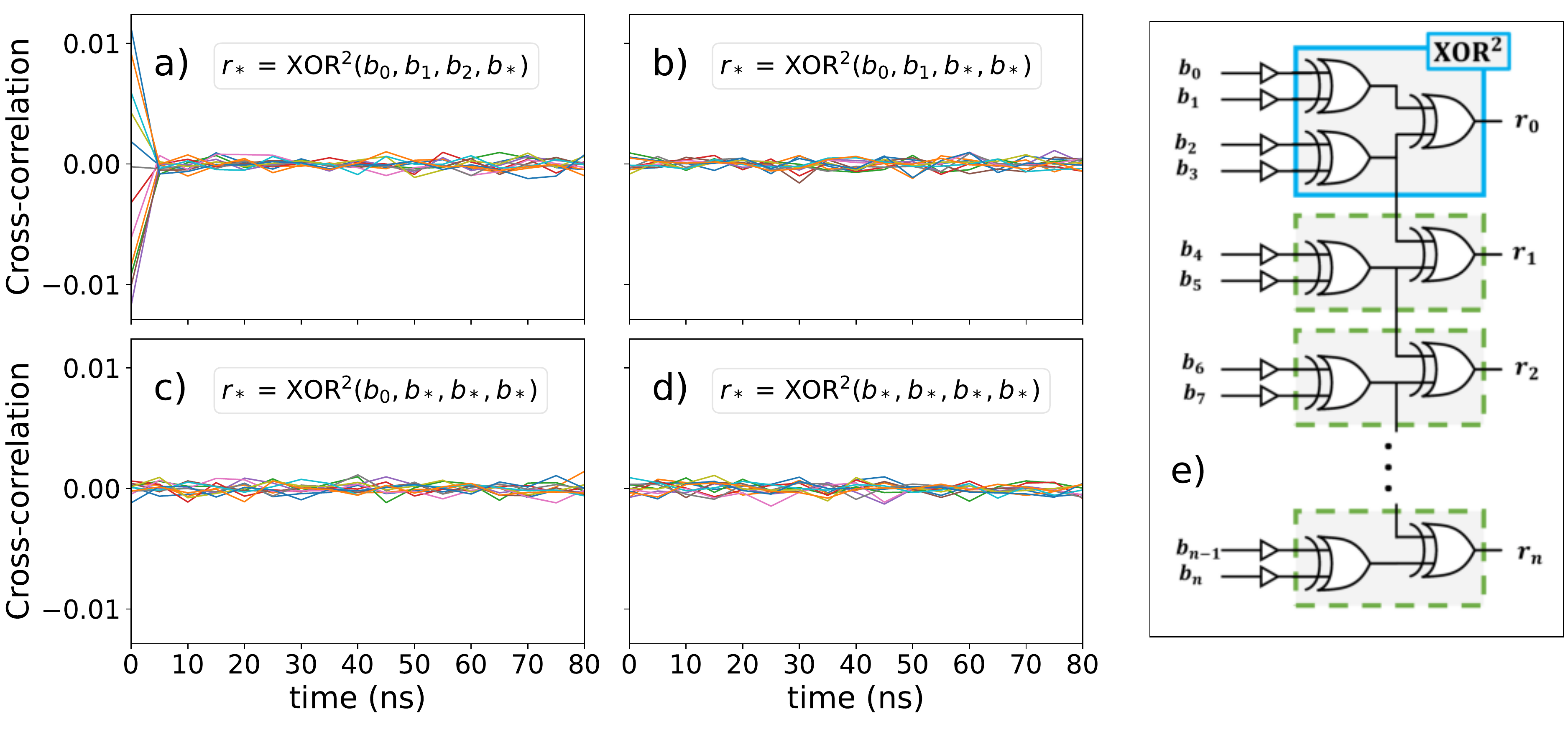}
\caption{\label{fig:XOR_circuit} a) - d) Cross-correlations of $r_0 = \textrm{XOR}^2(b_0,b_1,b_2,b_3)$ with XOR$^2(b_0,b_1,b_2,b_*)$, XOR$^2(b_0,b_1,b_*,b_*)$ and XOR$^2(b_*,b_*,b_*,b_*)$ for 20 different raw SMTJ bitstreams $b_*$. e) Raw bitstreams from SMTJs $b_i$ are combined with XOR gates to generate random bitstreams $r_i$. This circuit reduces the required number of SMTJs and XOR gates for generating multiple true random numbers.}
\end{figure}

Due to the intrinsic stochastic nature of SMTJs, the potential application as an energy-efficient true random number generator is of interest and has been proposed and studied by various groups \cite{RNG1, RNG2, RNG3, RNG4, RNG_rev1}.
To evaluate the quality of randomness of our stochastic MTJ signal and to clarify the use for potential cryptographic applications as a true random number generator, the National Institute of Standards and Technology Statistical Test Suite (NIST STS) \cite{NIST_STS} was applied to the binarized MTJ signal.
The amplified analog output signal was acquired by an oscilloscope and was converted into a Bernoulli sequence of 0s and 1s by comparing the output signal with the median of the acquired time series.
This ensures an equal probability of 0 and 1 of this Bernoulli sequence and is a prerequisite for a random bitstream.
The random bitstream can then be interpreted as the result of flipping an unbiased coin (head: 0, tail: 1) with a probability of 50$\,\%$ yielding to a head or a tail, where for true randomness the flips are independent of each other, such that any past result of a coin flip does not affect any future results.
The NIST Statistical Test Suite provides 15 different tests (188 tests including all subtests) and requires a minimum bitstream length of the order of $10^6$. 
There is a considerable amount of work that has been published recently regarding the quality of random number generation by superparamagnetic tunnel junctions \cite{ip_SMTJ_ns_Sun, RNG1, oop_SMTJ_us_Majetich, ip_SMTJ_us_Majetich}.
It has been shown that cryptographic quality randomness, by passing all NIST STS tests, is not reached by a single SMTJ bitstream but rather by combining multiple bitstreams with XOR gates (also called "XOR whitening").
This is a common method to improve the quality of randomness and is easy to implement by using standard CMOS-based exclusive or (XOR) gates.
To further improve the quality of randomness, nested XOR operations can be applied to multiple bitstreams.
In an XOR$^2$ operation, two rounds of XOR operations are applied to four input bitstreams (see blue rectangle in Fig.\ \ref{fig:XOR_circuit}e), whereas in an XOR$^3$ operation three rounds of XOR operations are applied to eight input bitstreams.
With this method the resulting bitstream of an XOR$^2$ operation passes all NIST STS tests, thus providing true randomness which can be used for cryptographic applications.
With an average auto-correlation time of $5\,$ns and four SMTJs as an input for an XOR$^2$ gate, true random bit generation of $200\,$Mbit/s can be achieved in a very energy-efficient way of only a few fJ/bit \cite{RNG1}.    
For the evaluation, $5\cdot 10^8$ data points were recorded, transformed into a binarized sequence and divided into eight bitstreams of equal size, in order to be able to feed the inputs to the XOR$^3$ gate.
Other works achieve cryptographic randomness by passing all NIST STS tests for a bitstream after a XOR$^2$ operation \cite{ip_SMTJ_ms_1} or XOR$^3$ operation \cite{RNG1}.
Here, we also observe true randomness after an XOR$^2$ operation for a sampling time of $5\,$ns, as shown in Table \ref{table:1}.
This sampling time matches the auto-correlation time measured for this time series.
Larger sampling times provide the same quality of randomness, whereas for sampling times below $5\,$ns each state would be sampled too frequently, thus resulting in non-true randomness with artificial and non-random chunks of only 0s and only 1s.\\
Randomness is a probabilistic property, therefore a random sequence can be evaluated in terms of probabilities of the occurrence of a specific pattern in the sequence by means of hypothesis tests.
The resulting p-values give the probabilities of obtaining the observed pattern assuming that the underlying sequence is random.
A common level of significance in cryptography for statistical hypothesis tests is $\alpha=0.01$, which was also chosen in this work.
For each of the 15 different tests of the NIST STS the p-value was calculated after applying different XOR operations to the binarized SMTJ telegraph signal.
All results are summarized in Table \ref{table:1}.
\begingroup
\begin{ruledtabular}
\begin{table}[H]
    \caption{Results of p-values for NIST STS tests after applying different XOR operations to binarized SMTJ telegraph signal. A randomness test is passed and highlighted in green for p-values above the significance level of 0.01. For multiple sub-tests, the particular p-values were combined and tested according to Fisher's method \cite{Fisher_method}.
    The sampling times of 3, 5 \& $10\,$ns are chosen to be close to $\tau_{ac}$.}
    \centering
    \small\addtolength{\tabcolsep}{-4pt}
    \begin{tabular}{c  c ccc ccc ccc}
         & Type & \multicolumn{3}{c}{Raw} & \multicolumn{3}{c}{XOR$^1$} & \multicolumn{3}{c}{XOR$^2$}\\
          & Sampling time & 3ns & 5ns & 10ns & 3ns & 5ns & 10ns & 3ns & 5ns & 10ns \\ \hline
           
 1.  &  Frequency   &0     &0     &0    &0  &\colorbox{lime}{0.141}     &0     &\colorbox{lime}{0.373}     &\colorbox{lime}{0.739}     &\colorbox{lime}{0.904}  \\  \hline 
 2.  &  Block  frequency  (m  =  128)   &0     &0     &0    &0 &0 &0 &0  &\colorbox{lime}{0.703}     &\colorbox{lime}{0.236}  \\  \hline 
 3.  &  Runs   &0     &0     &0     &0    &0  &0    &0  &\colorbox{lime}{0.069}     &\colorbox{lime}{0.817}  \\  \hline 
 4.  &  Longest-run   &0    &0 &0 &0  &\colorbox{lime}{0.019}     &0.001     &0.001     &\colorbox{lime}{0.990}     &\colorbox{lime}{0.185}  \\  \hline 
 5.  &  Binary  matrix  rank   &\colorbox{lime}{0.335}     &\colorbox{lime}{0.028}     &\colorbox{lime}{0.897}     &\colorbox{lime}{0.507}     &\colorbox{lime}{0.251}     &\colorbox{lime}{0.072}     &\colorbox{lime}{0.517}     &\colorbox{lime}{0.692}     &\colorbox{lime}{0.632}  \\  \hline 
 6.  &  Discrete  Fourier  transform  (spectral)   &0     &\colorbox{lime}{0.871}     &0    &0  &\colorbox{lime}{1}     &\colorbox{lime}{0.026}     &\colorbox{lime}{0.765}     &\colorbox{lime}{0.448}     &\colorbox{lime}{0.508}  \\  \hline 
 7.  &  Non-overlapping  template  matching   &0     &0     &0     &0    &0 &0 &0  &\colorbox{lime}{0.039}     &\colorbox{lime}{0.313}  \\  \hline 
 8.  &  Overlapping  template  matching   &0     &0    &0  &0    &0 &0 &0  &\colorbox{lime}{0.623}     &\colorbox{lime}{0.944}  \\  \hline 
 9.  &  Maurer’s  “universal  statistical”   &0     &0     &0    &0  &\colorbox{lime}{0.074}    &0  &\colorbox{lime}{0.182}     &\colorbox{lime}{0.472}     &\colorbox{lime}{0.761}  \\  \hline 
 10.  &  Linear  complexity  (M  =  500)   &\colorbox{lime}{0.467}     &\colorbox{lime}{0.137}     &\colorbox{lime}{0.737}     &\colorbox{lime}{0.967}     &\colorbox{lime}{0.929}     &\colorbox{lime}{0.986}     &\colorbox{lime}{0.174}     &\colorbox{lime}{0.085}     &\colorbox{lime}{0.673}  \\  \hline 
 11.  &  Serial  (m  =  16)   &0     &0     &0     &0    &0 &0 &0  &\colorbox{lime}{0.687}     &\colorbox{lime}{0.900}  \\  \hline 
 12.  &  Approximate  entropy  (m  =  10)  &0 &0 &0 &0 &0 &0 &0  &\colorbox{lime}{0.118}     &\colorbox{lime}{0.151}  \\  \hline 
 13.  &  Cumulative  sum  &0 &0 &0 &0 &0 &0  &\colorbox{lime}{0.162}     &\colorbox{lime}{0.172}     &\colorbox{lime}{0.010}  \\  \hline 
 14.  &  Random  excursion  &0  &\colorbox{lime}{0.027}     &\colorbox{lime}{0.068}     &\colorbox{lime}{0.068}     &\colorbox{lime}{0.073}     &\colorbox{lime}{0.042}     &\colorbox{lime}{0.073}     &\colorbox{lime}{0.074}     &\colorbox{lime}{0.074}  \\  \hline 
 15.  &  Random  excursion  variant  &0 &0 &0  &0     &\colorbox{lime}{0.015}    &0  &\colorbox{lime}{0.014}     &\colorbox{lime}{0.015}     &\colorbox{lime}{0.028}  \\  

    \end{tabular}
    \label{table:1}
\end{table}
\end{ruledtabular}
\endgroup
Raw time series data in our case never passes all NIST STS tests, independent of the chosen sampling rate.
If a p-value is greater than 0.01 (significance level) for a particular test then the bitstream is characterized as random and passes this test, whereas for a p-value below 0.01 the null hypothesis (of true randomness) is rejected and fails this test.

XOR operations can easily be implemented in CMOS and in our case, a simple random number generator would require 3 XOR gates and 4 SMTJs.
However, for further random number generators, less hardware is necessary since it is possible to tap the adjacent source randomness and XOR it with a new bitstream coming from the next two SMTJs.
This is depicted schematically in Fig.\ \ref{fig:XOR_circuit}e, where raw bitstreams from SMTJs $b_i$ are combined with XOR gates to generate random bitstreams $r_i$.
If $r_0$, which is generated by $b_0$ - $b_3$ is truly random, then $r_1$ - $r_n$ will provide true random numbers too, due to the same XOR$^2$ operation.
Nonetheless, an additional true random number only requires two more SMTJs and two more XOR gates (for $r_1$, $b_4$ and $b_5$ are added).
A further condition requires the true random numbers $r_0$ - $r_n$ to be uncorrelated.
Therefore we determined the cross-correlation of any new random number $r_*$ with $r_0$, as shown in Fig.\ \ref{fig:XOR_circuit}a-d.
Cross-correlations are calculated for 20 different bitstreams ($b_0$ - $b_{19}$) of length $5\cdot 10^6$.
Fig.\ \ref{fig:XOR_circuit}a points out a significant cross-correlation at 0 time lag for $r_0$ and $r_* = \textrm{XOR}^2(b_0,b_1,b_2,b_*)$, where $b_*$ is the added new SMTJ bitstream.
When two or more new SMTJs are combined, there is no further cross-correlation observed, meaning the output will be uncorrelated with $r_0$.
The sequence of SMTJ bitstreams to the XOR$^2$ circuit does not affect the outcome, since the XOR operation is permutation invariant.
As long as two new bitstreams ($b_*$ and $b_*$) are combined with two already used bitstreams ($b_2$ and $b_3$), it will result in a true random and uncorrelated number. 
It is also confirmed by the comparison with the cross-correlation in Fig.\ \ref{fig:XOR_circuit}d with $r_*=\textrm{XOR}^2(b_*,b_*,b_*,b_*)$, where no qualitative difference is recognizable.
Here, all SMTJs are exchanged, thus $r_*$ has to be uncorrelated with $r_0$.
Next, we evaluate how an applied MTJ current controls the bitstream generation rate and modifies the state probability.
Both features are important for the performance of a true random number generator.

\section{Joule heating and STT effects} 
\begin{figure}[H] 
\centering
\includegraphics[width=0.8\textwidth]{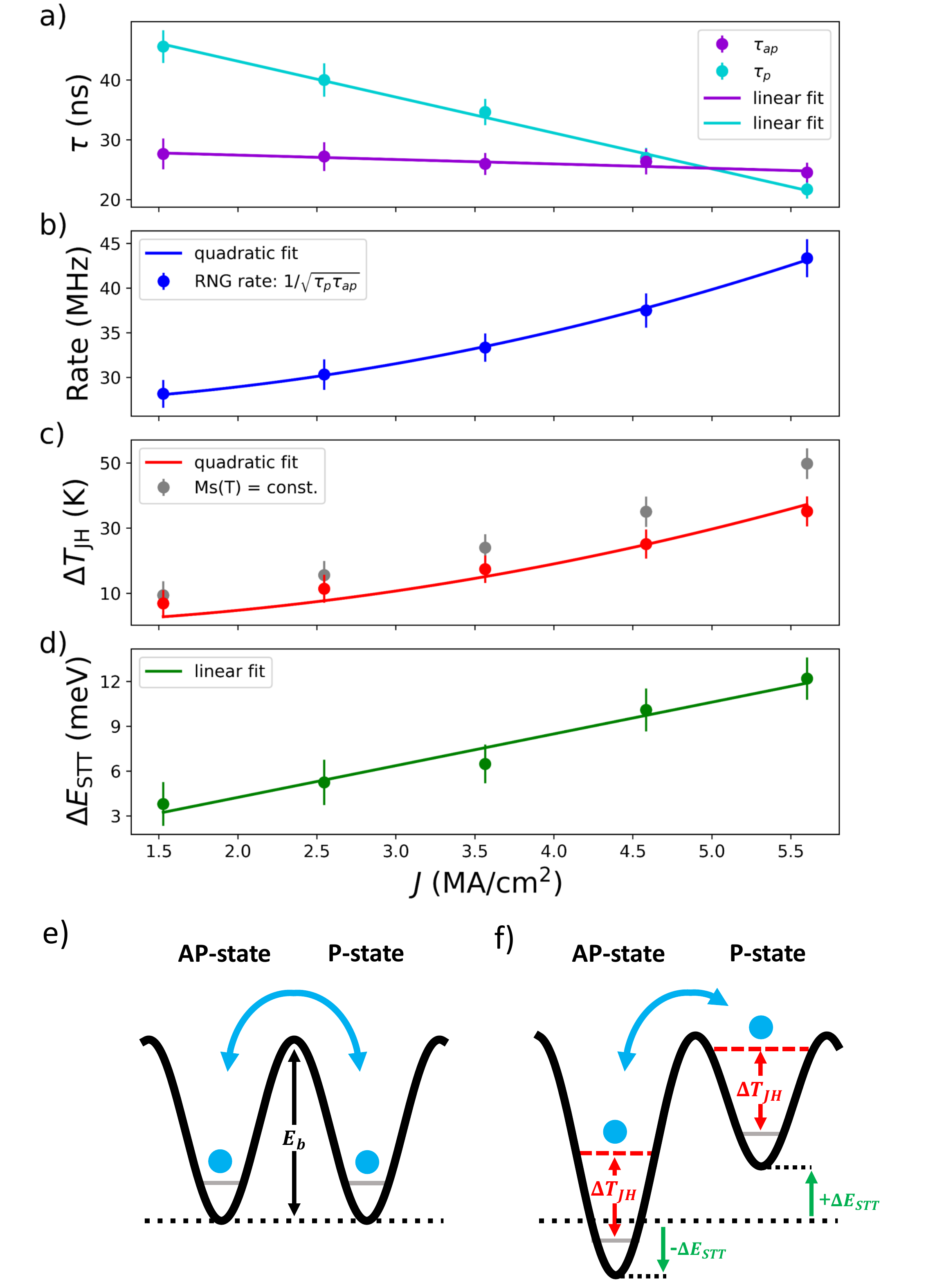}
\caption{\label{fig:Joule_heating}
a) SMTJ dwell times are plotted for current densities ranging from $1.5$ to $5.6\,$MA/cm$^2$. Linear fits are applied to indicate the trends. b) The average fluctuation rate is plotted in blue and fitted with a quadratic function. c) The temperature increase due to Joule heating is plotted in red. The non-linear increase is fitted with a quadratic function. The grey data points indicate the calculated temperature increase without the consideration of temperature dependent $M_s$. d) The energy shift due to the STT is plotted in green. Here, a positive energy shift destabilizes the P-state. e) A schematic of the free layer magnetization orientation described as a quasiparticle in a 1D symmetrical potential well, with two metastable states separated by an energy barrier. f) The effect of STT antisymmetrically shifts the energy level of P- and AP-state and the effect of Joule heating raises the energy level of the quasiparticle symmetrically for both states.}
\end{figure}

\begin{figure}[H]
\centering
\includegraphics[width=0.6\textwidth]{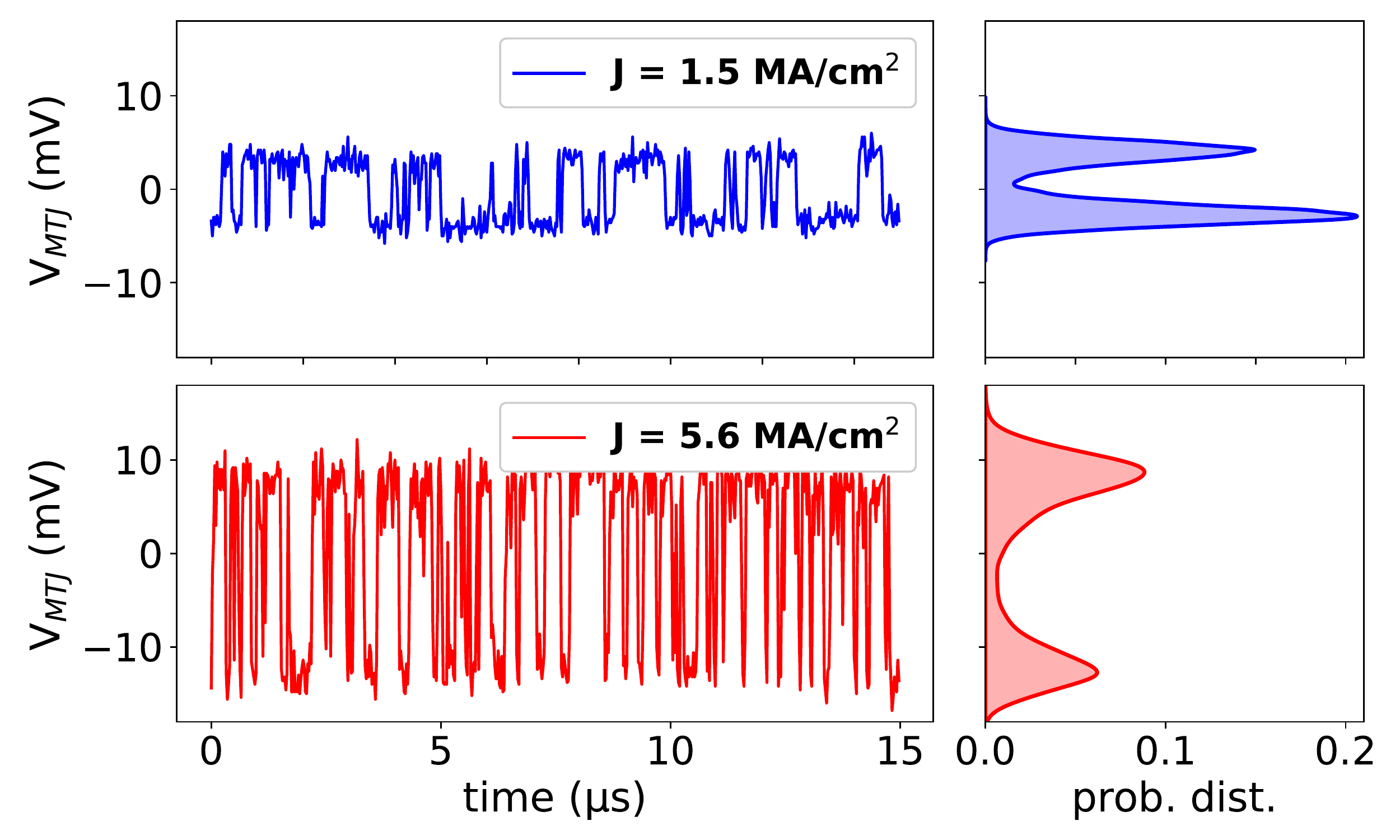}
\caption{\label{fig:time_series_low_high_J}Time series measurements of the amplified fluctuating MTJ signal at low ($1.5\,$MA/cm$^2$) and high ($5.6\,$MA/cm$^2$) current density together with the respective probability distribution of each signal.}
\end{figure}

Applying a current to an SMTJ can affect its switching speed, state probability and dwell times not only by STT, but also by Joule heating at the site of the free layer at the tunneling barrier.
Joule heating most often was studied in non-volatile MTJs, where a reduction of critical switching current is desired in order to perform fast and energy-efficient free layer switching \cite{MTJ_Joule_heating1,MTJ_Joule_heating2}.
The current through an MTJ for an applied voltage is mainly determined by the RA product and the areal size of the junction.
The RA product, which is strongly dependent on the thickness of the tunneling barrier, defines the current density of any nanopillar size because of the relation: $J=V/RA$.
For a RA product of $15\,\Omega$µm$^2$ and a few hundred millivolts, the typical current density is of the order of $1\,$MA/cm$^2$, where Joule heating can be significant.
A simulation for a $50\,$nm diameter MTJ suggests that the barrier temperature could increase more than $100\,$K after applying $8\,$MA/cm$^2$ for a few hundred nanoseconds \cite{MTJ_Joule_heating2}
and according to the Néel-Arrhenius law this will significantly decrease the dwell time of an SMTJ. 
Here, we have measured dwell times for different current densities and determined the contributions of STT and Joule heating. 
Figure \ref{fig:Joule_heating}e and \ref{fig:Joule_heating}f illustrate the effect of spin-transfer-torque and Joule heating.
Initially, without STT, both states (P- and AP-state) are tuned to the same energy level, where the free layer macrospin is considered as a quasiparticle in 1D potential well. 
Ambient thermal energy ($\approx 25\,$meV) raises the energy of the quasiparticle, such that the likelihood to overcome the energy barrier is enhanced.
By applying a current to the MTJ, the energy landscape is modified as illustrated in Fig.\ \ref{fig:Joule_heating}f, where the AP-state energy level is decreased and the P-state energy level is enhanced by STT in our circuit.
In the quasiparticle picture, this results in an energy shift $\Delta E_{STT}$ in the energy landscape and is here defined as a positive energy shift at the P-state (see Fig.\ \ref{fig:Joule_heating}f).
Higher current densities result in a higher energy level for the P-state yielding to shorter dwell times as it can be seen for $\tau_p$ in Fig.\ \ref{fig:Joule_heating}a.
The energy shift $\Delta E_{STT}$ is positive for lower current densities, which means that the AP-energy state is decreased whereas the P-energy state is increased.
Furthermore, Joule heating, originating from electron-phonon interaction, raises the energy level of the quasiparticle in the valleys symmetrically leading to overall faster switching.
The temperature increase with respect to room temperature $\Delta T_{JH}$ reveals a quadratic relation to $J$ (see Fig.\ \ref{fig:Joule_heating}c) as expected for heating power, which scales with $J^2$.
Fig.\ \ref{fig:Joule_heating}b reveals a similar trend of the average fluctuation rate, which is caused by the Joule heating effect.
This would correspond to a higher bit generation rate for a random number generator.
The maximum temperature increase at $5.6\,$MA/cm$^2$ reaches $36\pm5\,$K and corresponds to an applied voltage at the tunnel junction of approximately $0.6\,$V. 
The time series for the lowest and highest applied current densities are plotted in Fig.\ \ref{fig:time_series_low_high_J}.
Faster fluctuations with larger amplitude can be observed at high current densities, whereas for low current densities the fluctuation rate and amplitude are lower.
By this measurement, we demonstrate that Joule heating has a significant effect on the dwell times of an SMTJ and cannot be neglected.
For a potential application as a random number generator or as a p-bit for a probabilistic computer, the random number generation rate can be tuned by the current density.
When considering an SMTJ as a spiking neuron for neuromorphic computing \cite{Mizrahi1}, the current through the junction will modify the spike rate, thus enabling for instance rate coding in a neuromorphic network.

\subsection*{Calculation of $\Delta T_{JH}$ and $\Delta E_{STT}$}
The energy shift of the states $\Delta E_{STT}$ affects the dwell times according to:
\begin{equation}
\tau_{p,ap} = \tau_0 \exp \left(\frac{E_b \mp \Delta E_{STT}} {k_{\mathrm{B}} T} \right)
\end{equation}
where $\tau_{p,ap}$ are the dwell times, $T$ the temperature, $\tau_0$ the attempt time and $E_b$ the energy barrier between the metastable states.
The energy barrier $E_b$ for given dwell times can be determined by:
\begin{equation}
E_b = 1/2 k_{\mathrm{B}} T \log(\tau_p \tau_{ap} / \tau_0^2)
\label{eq:Eb}
\end{equation}
since the product of $\tau_p$ and $\tau_{ap}$ is independent of the energy barrier modification by STT. 
For this reason, the product of dwell times for an enhanced current density 
is also independent of STT.
\begin{equation}
\tau_{p}^* \tau_{ap}^* = \tau_0^2 \exp\left(\frac{2E_b^*}{k_{\mathrm{B}} T^*} \right)
\label{eq:epsilon}
\end{equation}
Here, $T^*$ and $E_b^*$ are considered as the altered (in our case increased) temperature and energy barrier with dwell times $\tau_p^*$ and $\tau_{ap}^*$.
The increased temperature can be described as $T^*=T+\Delta T_{JH}$ with $T=293\,$K.
The energy barrier at elevated temperature $E_b^*=E_b(T^*)=E_b(M_s(T^*))$ is a function of the temperature dependent saturation magnetization $M_s$ of the free layer Co$_{40}$Fe$_{40}$B$_{20}$, which approximately follows Bloch's $T^{3/2}$ law: $M_s(T)=M_s(0)(1-(T/T_c)^{3/2})$.
For the Curie temperature $T_c$ of Co$_{40}$Fe$_{40}$B$_{20}$ we have chosen $895\,$K \cite{Tc_of_cofeb}.
The in-plane crystalline anisotropy is considered to be constant since a temperature increase of a few Kelvin is unlikely to affect the polycrystalline phase of the free layer.
The resulting temperature dependent energy barrier $E_b^*$ is then described as:
\begin{equation}
E_b^*=E_b \frac{1-(T^*/T_c)^{3/2}}{1-(T/T_c)^{3/2}}
\label{eq:Eb_Bloch}
\end{equation}
The detailed equation between dwell times and temperature for the determination of $\Delta T_{JH}$  can be found in the supplementary information (Eq.\ \ref{eq:app:dT}).
The asymmetric contribution to the energy barrier, due to STT, is calculated by $\Delta E_{STT}=1/2 k_{\mathrm{B}} (T + \Delta T_{JH}) \log(\tau_{ap} / \tau_p)$.
However, one has to take into account that the initial P,AP-states at low current densities are not in equilibrium, because $\tau_p\neq \tau_{ap}$ (see Fig.\ \ref{fig:Joule_heating}a).
Shorter AP-dwell times arise due to an uncompensated stray field from the TMR stack itself and/or from an external applied in-plane field.
With linear fits to the dwell times, we can estimate the theoretical dwell time values at zero applied current $\tau_p(J=0)$, $\tau_{ap}(J=0)$, which is not measurable.
However, at $J=0$ there is neither a temperature increase ($\Delta T_{JH}=0$) nor an STT effect ($\Delta E_{STT}=0$) and the energy barrier at $293\,$K can be determined by equation \ref{eq:Eb} to be $95.3\pm1.1\,$meV.
Since the estimated dwell times at $J=0$ are not equal and the energy landscape is asymmetric at this point, this shows an initial energy shift of the states due to the Zeeman energy, which is here approximately $11\,$meV.
From equation \ref{eq:epsilon} the temperature increase can be calculated numerically (see also supplementary information S2 Eq.\ \ref{eq:app:dT}).\\ 

\section{Conclusion} 
In this work nanosecond superparamagnetic switching in in-plane MTJs with average dwell times of $6.7\pm0.1\,$ns at room temperature is demonstrated and the potential application as a true random bitstream generator is evaluated by the NIST Statistical Test Suite.
An XOR$^2$ operation on four independent SMTJ bitstreams results in a true random bit generation with a high throughput rate of $200\,$Mbit/s.
We then tune the stability and switching between the states and analyze the current dependence. We show in particular that the dwell times are linearly dependent on the applied current density and that Joule heating and spin-transfer-torque, both have a significant effect on dwell times as well as the dwell time ratio $\tau_p / \tau_{ap}$.
STT shifts the energy levels of the MTJ states antisymmetric and has a linear relation to the current density.
Joule heating raises both energy levels equally and has a quadratic dependence on the current density.
At the free layer, it can reach up to $329\pm5\,$K for $5.6\,$MA/cm$^2$ at ambient room temperature.
Since the dwell time is strongly dependent on temperature, shortest dwell times are observed for highest current densities.
This effect is not negligible and therefore should always be considered in superparamagnetic MTJs with low RA product.

\section*{Acknowledgments}
This work was supported by the Max Planck Graduate Center with the Johannes Gutenberg-Universität Mainz (MPGC) and used infrastructure provided by ForLab MagSens.
We acknowledge support by the Deutsche Forschungsgemeinschaft (DFG, German Research Foundation) project 268565370 (SFB TRR173 projects A01 and B02) as well as TopDyn and the Zeiss foundation through the Center for Emergent Algorithmic Intelligence and the Horizon Europe project no. 101070290 (NIMFEIA).
We would also like to thank T. Reimer for technical support during the development of the samples. 

\section*{Supplementary information}

S1: SMTJ voltage signal\\
$V_{src}:$ source voltage (DC), $V_{pp}:$ peak-to-peak voltage signal, $Z_i:$ input impedance of amplifier ($50\,\Omega$), $R_s:$ source resistance, $R_{MTJ}:$ MTJ resistance in either low (parallel) $R_P$ or high (anti-parallel) $R_{AP}$ configuration. 
\begin{equation}
\begin{aligned}
V &= V_{src} \frac{R_{MTJ}||Z_i}{R_s + R_{MTJ}||Z_i}
= \frac{ \frac{R_{MTJ}+Z_i}{R_{MTJ}Z_i}} {R_s + \frac{R_{MTJ}+Z_i}{R_{MTJ}Z_i}}\\
V_{pp} &= V_{AP} - V_{P} = V_{src} \left[ \frac{ \frac{R_{AP}+Z_i}{R_{AP}Z_i}} {R_s + \frac{R_{AP}+Z_i}{R_{AP}Z_i}} - \frac{ \frac{R_{P}+Z_i}{R_{P}Z_i}} {R_s + \frac{R_{P}+Z_i}{R_{P}Z_i}}  \right]
\end{aligned}
\label{eq:SMTJ_signal}
\end{equation}
For an MTJ resistance of $5-10\,$k$\Omega$, $R_s=5\,$k$\Omega$ and $V_{src}= 500\,$mV the fluctuating MTJ voltage signal ($V_{pp}$) is in the order of few mV. \\

S2: Derivation of the effect of Joule heating:\\
The product of dwell times at temperature $T$ is given by:
$\tau_p \tau_{ap} = \tau_0^2 \exp(2 E_b / k_{\mathrm{B}} T)$.
For an elevated temperature $T^*=T+\Delta T_{JH}$, the energy barrier modifies according to Bloch's $T^{3/2}$ law (see Eq. \ref{eq:Eb_Bloch}).
The following equation represents the dependence of Joule heating $\Delta T_{JH}$ on dwell times and can be solved numerically for $\Delta T_{JH}$.
\begin{equation}
\begin{aligned}
\tau_{p}^* \tau_{ap}^* &= \tau_0^2 \exp\left(\frac{2E_b^*}{k_{\mathrm{B}} T^*} \right) \\
\log \left( \frac{\tau_{p}^* \tau_{ap}^*}{\tau_0^2} \right) &= \frac{2E_b}{k_{\mathrm{B}} T^*} \cdot \frac{1-(T^*/T_c)^{3/2}}{1-(T/T_c)^{3/2}} \\
\log \left( \frac{\tau_{p}^* \tau_{ap}^*}{\tau_0^2} \right) &= \frac{T \log(\tau_p \tau_{ap} / \tau_0^2) (1-((T+ \Delta T_{JH})/T_c)^{3/2})}{ (T + \Delta T_{JH})(1-(T/T_c)^{3/2})} \\
\frac{\log(\tau_{p}^* \tau_{ap}^* / \tau_0^2)}{\log(\tau_p \tau_{ap} / \tau_0^2)} &= \frac{T (1-((T+ \Delta T_{JH})/T_c)^{3/2})}{ (T + \Delta T_{JH})(1-(T/T_c)^{3/2})}
\end{aligned}
\label{eq:app:dT}
\end{equation}
Here, $\tau_{p}^*$ and $\tau_{ap}^*$ refer to dwell times at an increased current density, while $\tau_{p}$ and $\tau_{ap}$ consider (reference) dwell times at a current density close to 0, where no Joule heating effect is present.

S3: Fisher's method  \cite{Fisher_method}:\\
Fisher's method is often used in meta-analysis to combine the results, namely p-values, of independent hypothesis test (each having the same hypothesis) to a new null hypothesis suggesting that all null hypothesis are true.
For a rejection H$_0$ at least one of the null hypothesis is rejected.
Since the p-value of a hypothesis test follows a uniform distribution on the interval [0,1], it can be shown that $-2 \sum_{i=1}^n \log(p_i)$ follows a chi-squared distribution $\chi^2_{2n}$ with 2$n$ degrees of freedom.
In this work, p-values of several sub-tests are combined and evaluated according to Fisher's method and the resulting p-value is noted in the table.
This holds for tests 11, 13, 14 and 15.\\

\section*{Data availability}
The data that support the findings of this study are available from the corresponding author upon request.
\bibliographystyle{naturemag}
\bibliography{main}

\begin{thebibliography}{10}
\expandafter\ifx\csname url\endcsname\relax
  \def\url#1{\texttt{#1}}\fi
\expandafter\ifx\csname urlprefix\endcsname\relax\def\urlprefix{URL }\fi
\providecommand{\bibinfo}[2]{#2}
\providecommand{\eprint}[2][]{\url{#2}}

\bibitem{MRAM_overview1}
\bibinfo{author}{Apalkov, D.}, \bibinfo{author}{Dieny, B.} \&
  \bibinfo{author}{Slaughter, J.~M.}
\newblock \bibinfo{title}{Magnetoresistive random access memory}.
\newblock \emph{\bibinfo{journal}{Proc. IEEE}} \textbf{\bibinfo{volume}{104}},
  \bibinfo{pages}{1796--1830} (\bibinfo{year}{2016}).

\bibitem{SOT_SMTJ1}
\bibinfo{author}{Ostwal, V.} \& \bibinfo{author}{Appenzeller, J.}
\newblock \bibinfo{title}{Spin--orbit torque-controlled magnetic tunnel
  junction with low thermal stability for tunable random number generation}.
\newblock \emph{\bibinfo{journal}{IEEE Magn. Lett.}}
  \textbf{\bibinfo{volume}{10}}, \bibinfo{pages}{1--5} (\bibinfo{year}{2019}).

\bibitem{neuro_spintronics_rev1}
\bibinfo{author}{Grollier, J.} \emph{et~al.}
\newblock \bibinfo{title}{Neuromorphic spintronics}.
\newblock \emph{\bibinfo{journal}{Nat. Electron.}}
  \textbf{\bibinfo{volume}{3}}, \bibinfo{pages}{360--370}
  (\bibinfo{year}{2020}).

\bibitem{Camsari_pbits}
\bibinfo{author}{Camsari, K.~Y.}, \bibinfo{author}{Faria, R.},
  \bibinfo{author}{Sutton, B.~M.} \& \bibinfo{author}{Datta, S.}
\newblock \bibinfo{title}{Stochastic p-bits for invertible logic}.
\newblock \emph{\bibinfo{journal}{Phys. Rev. X}} \textbf{\bibinfo{volume}{7}},
  \bibinfo{pages}{031014} (\bibinfo{year}{2017}).

\bibitem{BM_pbits}
\bibinfo{author}{Kaiser, J.} \emph{et~al.}
\newblock \bibinfo{title}{Hardware-aware in situ learning based on stochastic
  magnetic tunnel junctions}.
\newblock \emph{\bibinfo{journal}{Phys. Rev. Appl.}}
  \textbf{\bibinfo{volume}{17}}, \bibinfo{pages}{014016}
  (\bibinfo{year}{2022}).

\bibitem{RC_pbits}
\bibinfo{author}{Ganguly, S.}, \bibinfo{author}{Camsari, K.~Y.} \&
  \bibinfo{author}{Ghosh, A.~W.}
\newblock \bibinfo{title}{Reservoir computing using stochastic p-bits}.
\newblock \emph{\bibinfo{journal}{arXiv:1709.10211}}  (\bibinfo{year}{2017}).

\bibitem{spiking_NN1}
\bibinfo{author}{Sengupta, A.}, \bibinfo{author}{Panda, P.},
  \bibinfo{author}{Wijesinghe, P.}, \bibinfo{author}{Kim, Y.} \&
  \bibinfo{author}{Roy, K.}
\newblock \bibinfo{title}{Magnetic tunnel junction mimics stochastic cortical
  spiking neurons}.
\newblock \emph{\bibinfo{journal}{Sci. Rep.}} \textbf{\bibinfo{volume}{6}},
  \bibinfo{pages}{1--8} (\bibinfo{year}{2016}).

\bibitem{spiking_NN2}
\bibinfo{author}{Sengupta, A.}, \bibinfo{author}{Parsa, M.},
  \bibinfo{author}{Han, B.} \& \bibinfo{author}{Roy, K.}
\newblock \bibinfo{title}{Probabilistic deep spiking neural systems enabled by
  magnetic tunnel junction}.
\newblock \emph{\bibinfo{journal}{IEEE Trans. Electron Devices}}
  \textbf{\bibinfo{volume}{63}}, \bibinfo{pages}{2963--2970}
  (\bibinfo{year}{2016}).

\bibitem{SC_Daniels}
\bibinfo{author}{Daniels, M.~W.}, \bibinfo{author}{Madhavan, A.},
  \bibinfo{author}{Talatchian, P.}, \bibinfo{author}{Mizrahi, A.} \&
  \bibinfo{author}{Stiles, M.~D.}
\newblock \bibinfo{title}{Energy-efficient stochastic computing with
  superparamagnetic tunnel junctions}.
\newblock \emph{\bibinfo{journal}{Phys. Rev. Appl.}}
  \textbf{\bibinfo{volume}{13}}, \bibinfo{pages}{034016}
  (\bibinfo{year}{2020}).

\bibitem{MCMC_book}
\bibinfo{author}{Gilks, W.~R.}, \bibinfo{author}{Richardson, S.} \&
  \bibinfo{author}{Spiegelhalter, D.}
\newblock \emph{\bibinfo{title}{Markov chain Monte Carlo in practice}}
  (\bibinfo{publisher}{CRC press}, \bibinfo{year}{1995}).

\bibitem{Camsari_traveling_salesman}
\bibinfo{author}{Camsari, K.~Y.}, \bibinfo{author}{Sutton, B.~M.} \&
  \bibinfo{author}{Datta, S.}
\newblock \bibinfo{title}{P-bits for probabilistic spin logic}.
\newblock \emph{\bibinfo{journal}{Appl. Phys. Rev.}}
  \textbf{\bibinfo{volume}{6}}, \bibinfo{pages}{011305} (\bibinfo{year}{2019}).

\bibitem{ip_SMTJ_ms_1}
\bibinfo{author}{Kim, T.} \emph{et~al.}
\newblock \bibinfo{title}{Demonstration of in-plane magnetized stochastic
  magnetic tunnel junction for binary stochastic neuron}.
\newblock \emph{\bibinfo{journal}{AIP Adv.}} \textbf{\bibinfo{volume}{12}},
  \bibinfo{pages}{075104} (\bibinfo{year}{2022}).

\bibitem{ip_SMTJ_us_1}
\bibinfo{author}{Zink, B.~R.}, \bibinfo{author}{Lv, Y.} \&
  \bibinfo{author}{Wang, J.-P.}
\newblock \bibinfo{title}{Telegraphic switching signals by magnet tunnel
  junctions for neural spiking signals with high information capacity}.
\newblock \emph{\bibinfo{journal}{J. Appl. Phys.}}
  \textbf{\bibinfo{volume}{124}}, \bibinfo{pages}{152121}
  (\bibinfo{year}{2018}).

\bibitem{ip_SMTJ_us_Majetich}
\bibinfo{author}{Bapna, M.} \& \bibinfo{author}{Majetich, S.~A.}
\newblock \bibinfo{title}{Current control of time-averaged magnetization in
  superparamagnetic tunnel junctions}.
\newblock \emph{\bibinfo{journal}{Appl. Phys. Lett.}}
  \textbf{\bibinfo{volume}{111}}, \bibinfo{pages}{243107}
  (\bibinfo{year}{2017}).

\bibitem{ip_SMTJ_us_Majetich2}
\bibinfo{author}{Parks, B.}, \bibinfo{author}{Abdelgawad, A.},
  \bibinfo{author}{Wong, T.}, \bibinfo{author}{Evans, R.~F.} \&
  \bibinfo{author}{Majetich, S.~A.}
\newblock \bibinfo{title}{Magnetoresistance dynamics in superparamagnetic co-
  fe- b nanodots}.
\newblock \emph{\bibinfo{journal}{Phys. Rev. Appl.}}
  \textbf{\bibinfo{volume}{13}}, \bibinfo{pages}{014063}
  (\bibinfo{year}{2020}).

\bibitem{ip_SMTJ_ns_Sun}
\bibinfo{author}{Safranski, C.} \emph{et~al.}
\newblock \bibinfo{title}{Demonstration of nanosecond operation in stochastic
  magnetic tunnel junctions}.
\newblock \emph{\bibinfo{journal}{Nano Lett.}} \textbf{\bibinfo{volume}{21}},
  \bibinfo{pages}{2040--2045} (\bibinfo{year}{2021}).

\bibitem{ip_SMTJ_ns_Fukami}
\bibinfo{author}{Hayakawa, K.} \emph{et~al.}
\newblock \bibinfo{title}{Nanosecond random telegraph noise in in-plane
  magnetic tunnel junctions}.
\newblock \emph{\bibinfo{journal}{Phys. Rev. Lett.}}
  \textbf{\bibinfo{volume}{126}}, \bibinfo{pages}{117202}
  (\bibinfo{year}{2021}).

\bibitem{theory_relaxation_times}
\bibinfo{author}{Kanai, S.}, \bibinfo{author}{Hayakawa, K.},
  \bibinfo{author}{Ohno, H.} \& \bibinfo{author}{Fukami, S.}
\newblock \bibinfo{title}{Theory of relaxation time of stochastic nanomagnets}.
\newblock \emph{\bibinfo{journal}{Phys. Rev. B}}
  \textbf{\bibinfo{volume}{103}}, \bibinfo{pages}{094423}
  (\bibinfo{year}{2021}).

\bibitem{oop_SMTJ_ms_Reiss}
\bibinfo{author}{Reiss, G.}, \bibinfo{author}{Ludwig, J.} \&
  \bibinfo{author}{Rott, K.}
\newblock \bibinfo{title}{Superparamagnetic dwell times and tuning of switching
  rates in perpendicular {CoFeB/MgO/CoFeB} tunnel junctions}.
\newblock \emph{\bibinfo{journal}{arXiv:1908.02139}}  (\bibinfo{year}{2019}).

\bibitem{oop_SMTJ_ms_Majetich}
\bibinfo{author}{Bapna, M.} \emph{et~al.}
\newblock \bibinfo{title}{Magnetostatic effects on switching in small magnetic
  tunnel junctions}.
\newblock \emph{\bibinfo{journal}{Appl. Phys. Lett.}}
  \textbf{\bibinfo{volume}{108}}, \bibinfo{pages}{022406}
  (\bibinfo{year}{2016}).

\bibitem{Camsari_inter_fac}
\bibinfo{author}{Borders, W.~A.} \emph{et~al.}
\newblock \bibinfo{title}{Integer factorization using stochastic magnetic
  tunnel junctions}.
\newblock \emph{\bibinfo{journal}{Nature}} \textbf{\bibinfo{volume}{573}},
  \bibinfo{pages}{390--393} (\bibinfo{year}{2019}).

\bibitem{oop_SMTJ_us_Majetich}
\bibinfo{author}{Parks, B.} \emph{et~al.}
\newblock \bibinfo{title}{Superparamagnetic perpendicular magnetic tunnel
  junctions for true random number generators}.
\newblock \emph{\bibinfo{journal}{AIP Adv.}} \textbf{\bibinfo{volume}{8}},
  \bibinfo{pages}{055903} (\bibinfo{year}{2018}).

\bibitem{oop_SMTJ_us_Fukami}
\bibinfo{author}{Kobayashi, K.} \emph{et~al.}
\newblock \bibinfo{title}{Sigmoidal curves of stochastic magnetic tunnel
  junctions with perpendicular easy axis}.
\newblock \emph{\bibinfo{journal}{Appl. Phys. Lett.}}
  \textbf{\bibinfo{volume}{119}}, \bibinfo{pages}{132406}
  (\bibinfo{year}{2021}).

\bibitem{Joule_heating_Krause}
\bibinfo{author}{Krause, S.}, \bibinfo{author}{Herzog, G.},
  \bibinfo{author}{Schlenhoff, A.}, \bibinfo{author}{Sonntag, A.} \&
  \bibinfo{author}{Wiesendanger, R.}
\newblock \bibinfo{title}{Joule heating and spin-transfer torque investigated
  on the atomic scale using a spin-polarized scanning tunneling microscope}.
\newblock \emph{\bibinfo{journal}{Phys. Rev. Lett.}}
  \textbf{\bibinfo{volume}{107}}, \bibinfo{pages}{186601}
  (\bibinfo{year}{2011}).

\bibitem{NIST_STS}
\bibinfo{author}{Bassham~III, L.~E.} \emph{et~al.}
\newblock \emph{\bibinfo{title}{Sp 800-22 rev. 1a. a statistical test suite for
  random and pseudorandom number generators for cryptographic applications}}
  (\bibinfo{publisher}{National Institute of Standards \& Technology},
  \bibinfo{year}{2010}).

\bibitem{LS_ADI_paper}
\bibinfo{author}{Schnitzspan, L.} \emph{et~al.}
\newblock \bibinfo{title}{Impact of annealing temperature on tunneling
  magnetoresistance multilayer stacks}.
\newblock \emph{\bibinfo{journal}{IEEE Magn. Lett.}}
  \textbf{\bibinfo{volume}{11}}, \bibinfo{pages}{1--5} (\bibinfo{year}{2020}).

\bibitem{theory_exchange_bias}
\bibinfo{author}{Stamps, R.}
\newblock \bibinfo{title}{Mechanisms for exchange bias}.
\newblock \emph{\bibinfo{journal}{J. Phys. D. Appl. Phys.}}
  \textbf{\bibinfo{volume}{33}}, \bibinfo{pages}{R247} (\bibinfo{year}{2000}).

\bibitem{strayfield_pMTJ}
\bibinfo{author}{Jenkins, S.} \emph{et~al.}
\newblock \bibinfo{title}{Magnetic stray fields in nanoscale magnetic tunnel
  junctions}.
\newblock \emph{\bibinfo{journal}{J. Phys. D. Appl. Phys.}}
  \textbf{\bibinfo{volume}{53}}, \bibinfo{pages}{044001}
  (\bibinfo{year}{2019}).

\bibitem{cipt_tech}
\bibinfo{author}{Worledge, D.} \& \bibinfo{author}{Trouilloud, P.}
\newblock \bibinfo{title}{Magnetoresistance measurement of unpatterned magnetic
  tunnel junction wafers by current-in-plane tunneling}.
\newblock \emph{\bibinfo{journal}{Appl. Phys. Lett.}}
  \textbf{\bibinfo{volume}{83}}, \bibinfo{pages}{84--86}
  (\bibinfo{year}{2003}).

\bibitem{theory_Neel_law}
\bibinfo{author}{H{\"a}nggi, P.}, \bibinfo{author}{Talkner, P.} \&
  \bibinfo{author}{Borkovec, M.}
\newblock \bibinfo{title}{Reaction-rate theory: fifty years after {Kramers}}.
\newblock \emph{\bibinfo{journal}{Rev. Mod. Phys.}}
  \textbf{\bibinfo{volume}{62}}, \bibinfo{pages}{251} (\bibinfo{year}{1990}).

\bibitem{theory_low_barrier_m_IEEE}
\bibinfo{author}{Hassan, O.}, \bibinfo{author}{Faria, R.},
  \bibinfo{author}{Camsari, K.~Y.}, \bibinfo{author}{Sun, J.~Z.} \&
  \bibinfo{author}{Datta, S.}
\newblock \bibinfo{title}{Low-barrier magnet design for efficient hardware
  binary stochastic neurons}.
\newblock \emph{\bibinfo{journal}{IEEE Magn. Lett.}}
  \textbf{\bibinfo{volume}{10}}, \bibinfo{pages}{1--5} (\bibinfo{year}{2019}).

\bibitem{RNG1}
\bibinfo{author}{Vodenicarevic, D.} \emph{et~al.}
\newblock \bibinfo{title}{Low-energy truly random number generation with
  superparamagnetic tunnel junctions for unconventional computing}.
\newblock \emph{\bibinfo{journal}{Phys. Rev. Appl.}}
  \textbf{\bibinfo{volume}{8}}, \bibinfo{pages}{054045} (\bibinfo{year}{2017}).

\bibitem{RNG2}
\bibinfo{author}{Fukushima, A.} \emph{et~al.}
\newblock \bibinfo{title}{Spin dice: A scalable truly random number generator
  based on spintronics}.
\newblock \emph{\bibinfo{journal}{Appl. Phys. Express}}
  \textbf{\bibinfo{volume}{7}}, \bibinfo{pages}{083001} (\bibinfo{year}{2014}).

\bibitem{RNG3}
\bibinfo{author}{Chen, X.}, \bibinfo{author}{Zhang, J.}, \bibinfo{author}{Xiao,
  J.} \emph{et~al.}
\newblock \bibinfo{title}{Magnetic-tunnel-junction-based true random-number
  generator with enhanced generation rate}.
\newblock \emph{\bibinfo{journal}{Phys. Rev. Appl.}}
  \textbf{\bibinfo{volume}{18}}, \bibinfo{pages}{L021002}
  (\bibinfo{year}{2022}).

\bibitem{RNG4}
\bibinfo{author}{Rangarajan, N.}, \bibinfo{author}{Parthasarathy, A.} \&
  \bibinfo{author}{Rakheja, S.}
\newblock \bibinfo{title}{A spin-based true random number generator exploiting
  the stochastic precessional switching of nanomagnets}.
\newblock \emph{\bibinfo{journal}{J. Appl. Phys.}}
  \textbf{\bibinfo{volume}{121}}, \bibinfo{pages}{223905}
  (\bibinfo{year}{2017}).

\bibitem{RNG_rev1}
\bibinfo{author}{Fu, Z.} \emph{et~al.}
\newblock \bibinfo{title}{An overview of spintronic true random number
  generator}.
\newblock \emph{\bibinfo{journal}{Front. Phys.}} \bibinfo{pages}{172}
  (\bibinfo{year}{2021}).

\bibitem{Fisher_method}
\bibinfo{author}{Mosteller, F.} \& \bibinfo{author}{Fisher, R.~A.}
\newblock \bibinfo{title}{Questions and answers}.
\newblock \emph{\bibinfo{journal}{The American Statistician}}
  \textbf{\bibinfo{volume}{2}}, \bibinfo{pages}{30--31} (\bibinfo{year}{1948}).

\bibitem{MTJ_Joule_heating1}
\bibinfo{author}{Lee, D.} \& \bibinfo{author}{Lim, S.~H.}
\newblock \bibinfo{title}{Increase of temperature due to joule heating during
  current-induced magnetization switching of an mgo-based magnetic tunnel
  junction}.
\newblock \emph{\bibinfo{journal}{Appl. Phys. Lett.}}
  \textbf{\bibinfo{volume}{92}}, \bibinfo{pages}{233502}
  (\bibinfo{year}{2008}).

\bibitem{MTJ_Joule_heating2}
\bibinfo{author}{Kan, J.~J.} \emph{et~al.}
\newblock \bibinfo{title}{A study on practically unlimited endurance of
  stt-mram}.
\newblock \emph{\bibinfo{journal}{IEEE Trans. on Electron Devices}}
  \textbf{\bibinfo{volume}{64}}, \bibinfo{pages}{3639--3646}
  (\bibinfo{year}{2017}).

\bibitem{Mizrahi1}
\bibinfo{author}{Mizrahi, A.} \emph{et~al.}
\newblock \bibinfo{title}{Neural-like computing with populations of
  superparamagnetic basis functions}.
\newblock \emph{\bibinfo{journal}{Nat. Commun.}} \textbf{\bibinfo{volume}{9}},
  \bibinfo{pages}{1--11} (\bibinfo{year}{2018}).

\bibitem{Tc_of_cofeb}
\bibinfo{author}{Lee, K.-M.}, \bibinfo{author}{Choi, J.~W.},
  \bibinfo{author}{Sok, J.} \& \bibinfo{author}{Min, B.-C.}
\newblock \bibinfo{title}{Temperature dependence of the interfacial magnetic
  anisotropy in {W/CoFeB/MgO}}.
\newblock \emph{\bibinfo{journal}{AIP Adv.}} \textbf{\bibinfo{volume}{7}},
  \bibinfo{pages}{065107} (\bibinfo{year}{2017}).

\end{thebibliography}

\end{document}